\documentclass[%
 reprint,
 amsmath,amssymb,
 aps,
]{revtex4-1}

\usepackage{graphicx}
\usepackage{dcolumn}
\usepackage{bm}
\usepackage{color}

\begin{document}


\title{Splitting of elliptic flow in non-central relativistic heavy-ion collisions}

\author{Zhengyu Chen$^{1}$, Zeyan Wang$^{1}$, Carsten Greiner$^{2}$, Zhe Xu$^{1}$}
\affiliation{$^1$ Department of Physics, Tsinghua University and Collaborative Innovation Center of Quantum Matter, Beijing 100084, China}
\affiliation{$^2$ Institut f$\ddot{u}$r Theoretische Physik, Johann Wolfgang Goethe-Universit$\ddot{a}$t Frankfurt, Max-von-Laue-Strasse 1, 60438 Frankfurt am Main, Germany}

\date{\today}

\begin{abstract}
We predict a new effect due to the presence of the global vorticity in non-central relativistic
heavy-ion collisions, namely a splitting of the elliptic flow parameter $v_2$ at non-zero rapidity.
The size of the splitting is proposed as a new observable that can be used to constrain 
the initial vortical configuration of the produced QCD matter in experiments. The new
findings are demonstrated by numerical calculations employing 
the parton cascade model, Boltzmann Approach of MultiParton Scatterings (BAMPS), 
for non-central Au + Au collisions at $\sqrt{s_{NN}} = 200 \ GeV$.
\end{abstract}

\maketitle


Global spin polarization of hadrons, observed in non-central relativistic heavy-ion collisions
at the BNL Relativistic Heavy Ion Collider (RHIC) and the CERN Large Hadron Collider (LHC)
 \cite{STAR:2017ckg,Adam:2018ivw,Abelev:2007zk,Abelev:2008ag,Acharya:2019vpe},
indicates the generation of  a strong vorticity field. The magnitude of the global vorticity
is estimated to be $\omega \approx 10^{22} s^{-1}$ \cite{STAR:2017ckg}, which is
the highest value known in nature.
Such strong vorticity opens a new window for the study of the quark-gluon plasma (QGP)
in heavy-ion collisions. Theoretical developments such as spin kinetic theory 
\cite{Yang:2018lew,Mueller:2019gjj,Weickgenannt:2019dks,Gao:2019znl,Hattori:2019ahi,Wang:2019moi,Li:2019qkf,Yang:2020hri,Liu:2020flb,Bhadury:2020puc}
and spin hydrodynamics \cite{Florkowski:2017ruc,Florkowski:2018ahw,Montenegro:2018bcf,Hattori:2019lfp,Fukushima:2020ucl,Li:2020eon}
become new interdisciplinary research areas. 

Current studies are focusing on the initial vortical configuration of the QCD matter
in non-central heavy-ion collisions. The directed flow $v_1$ of the light charged hadrons 
\cite{STAR:2014clz,Bozek:2010bi,Nara:2016phs} and heavy flavor hadrons 
\cite{STAR:2019clv,Chatterjee:2017ahy,Chen:2019qzx,Oliva:2020doe} 
are suggested as probes to the initial tilted shape of matter. 
Still, there exists a spin sign problem \cite{Becattini:2020ngo,Liu:2020ymh,Gao:2020vbh}
between the experimental and theoretical results on the local polarization such as azimulthal
angle dependence, which demands deeper understanding to the initial vortical configuration.
In this Letter, we predict a new effect of the global vorticity and propose a new observable,
which can be used to constrain the initial vortical configuration of matter.

\begin{figure}
\includegraphics[width=8cm,height=6cm]{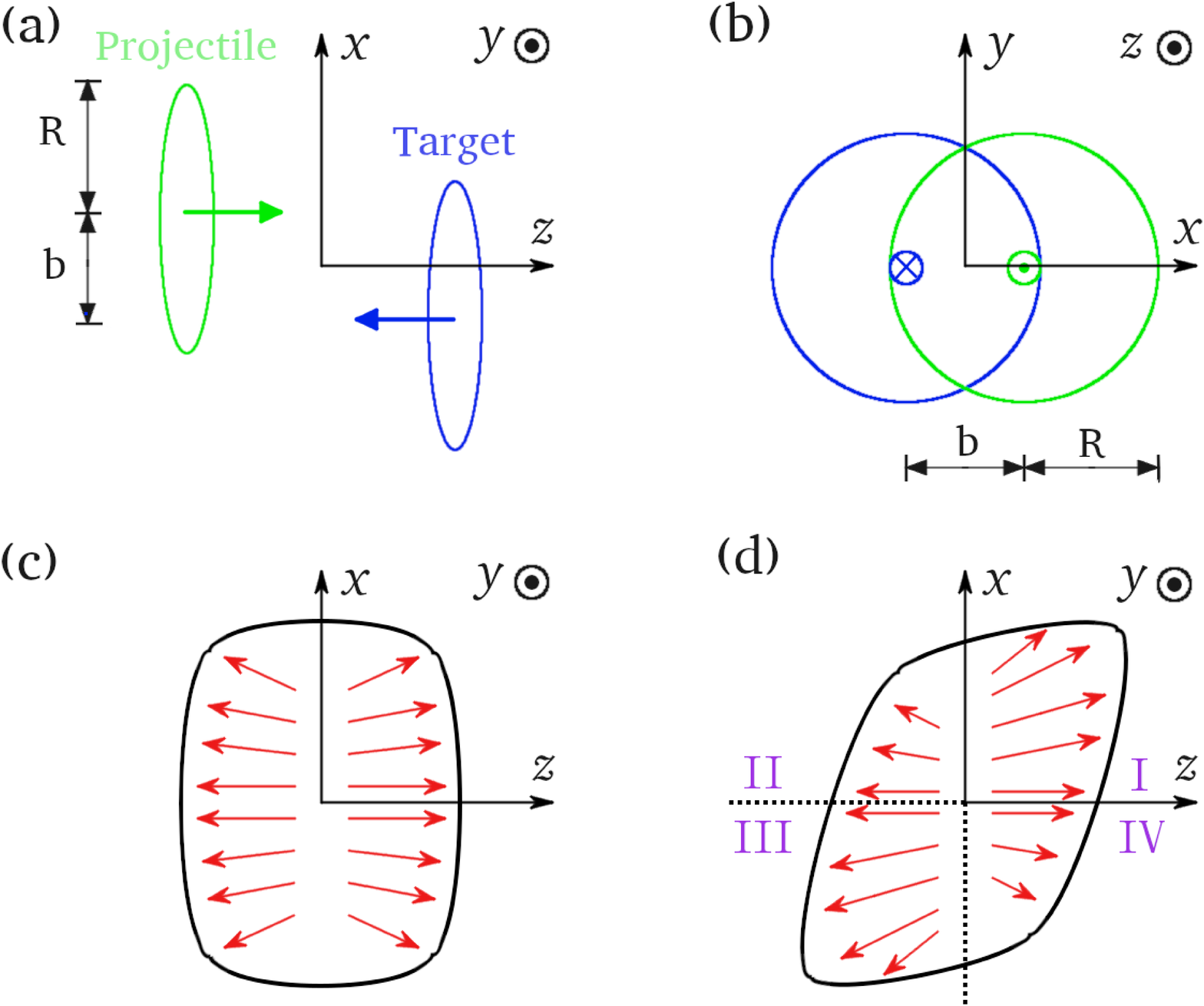}
\caption{\label{fig:config} A sketch of the collective motion of the QCD matter 
produced in a non-central relativistic heavy-ion collision.}
\end{figure}

The idea comes from the thought of a possible effect of the vortical motion on
the expansion transversely to the beam axis. For a rigid rotating plasma,  the authors in 
Ref.~\cite{Becattini:2007sr} found an enhancement of the elliptic flow $v_2$ due to
the centrifugal effect. However,  since the equation of state (EOS) of QGP is very soft,
the global orbital angular momentum would probably never result in a rigid rotation of
the produced QCD matter. Instead, it will give rise to the shear of the longitudinal flow,
which leads to a collective motion round the $y$ axis in addition to the
longitudinal expansion along the beam ($z$) axis and the transverse expansion
in the $x-y$ plane. The shape of the expanding matter in the reaction plane is sketched
in Fig. \ref{fig:config} ($d$), compared to the case without the vortical behavior
[see Fig. \ref{fig:config} ($c$)]. It seems that the vortical motion will ``drive'' the system
outwards in the I and III quadrant and ``press'' the system inwards in the II and IV quadrant.
Thus, the vortical motion breaks the mirror symmetry with regards to the $y-z$ plane
and also the Bjorken boost invariance, which hold in the case without the vortical motion.
Considering a particular piece at a space-time rapidity $\eta_s$, namely the medium
in the $x-y$ plane at a certain $\eta_s=1/2 \ln (t+z)/(t-z)$. Due to the symmetry breaking
at non-zero $\eta_s$,
the collective expansion in the half $x-y$ plane with positive $x$ is different from that
in the another half plane with negative $x$. This difference may be transferred into
the final particle momentum distribution. From this thought we predict that the elliptic
flow parameter $v_2$ at non-zero momentum rapidity measured using particles with
positive $p_x$ is different from that $v_2$ measured using particles with negative
$p_x$. We call this the splitting of the elliptic flow. In the following we will demonstrate 
the new finding by calculations within kinetic theory.

We employ the Monte Carlo kinetic transport model,  Boltzmann Approach of MultiParton
Scatterings (BAMPS)\cite{Xu:2004mz,Xu:2007aa}, to calculate the space-time evolution
of the quark gluon matter in heavy-ion collisions. BAMPS had successfully decribed the 
experimental data of elliptic flow  measured at RHIC and LHC
\cite{Xu:2007jv,Xu:2008av,Uphoff:2014cba}.

Now we briefly present the initial condition for subsequent kinetic transport
calculations shown later in this letter.
In the Glauber picture of heavy-ion collisions, quarks and gluons are initially produced
in ``hard'' binary nucleon-nucleon collisions as well as in ``soft'' collective collisions
between participant nucleons of the projectile and target nucleus. The latter can be
described in the wounded nucleon model  \cite{Miller:2007ri},  in which the number
of produced quarks and gluons is assumed to be proportional to the number of
participant nucleons and the quarks and gluons will take a part of the momentum of
the participant projectile and target nucleons. Due to the unequal local
number densities of participant projectile and target nucleons, the quark gluon system
possesses a global angular momentum round the $y$ axis. 

The participant nucleon number distribution in the transverse plane can be evaluated as
 \cite{Miller:2007ri}
\begin{eqnarray}
\frac{d N_{part}^{P,T}}{d{\bf x}_T} &=& T^{P,T}({\bf x}_T, b) \{ 1-exp [ -\sigma_{p+p} T^{T,P}({\bf x}_T, b) ] \} \,, 
\label{eq:f_eq}
\end{eqnarray}
where the superscript $P$ or $T$ denotes projectile or target, $\sigma_{p+p}$ is 
the nucleon-nucleon reaction cross section, which is approximately $42$ $mb$ for Au+Au
collisions at top RHIC energy, and  
\begin{equation}
T^{P,T}({\bf x}_T, b) = \int dz \, n^{P,T}_{WS}({\bf x}_T, z, b) 
\end{equation}
is the thickness function of the projectile (target) nucleus. Here $n^{P,T}_{WS}({\bf x}_T, z, b)$
is the Woods-Saxon distribution for the nuclear density of the colliding nucleus.
The participant nucleon number relative asymmetry distribution is defined as
\begin{equation}
\label{participant_asymmetry}
A_{part}({\bf x}_T, b) = \frac{dN_{part}^P/d{\bf x}_T-dN_{part}^T/d{\bf x}_T}{dN_{part}^P/d{\bf x}_T+dN_{part}^T/d{\bf x}_T}  \,.
\end{equation}

\begin{figure}[b]
\includegraphics[width=8cm,height=6cm]{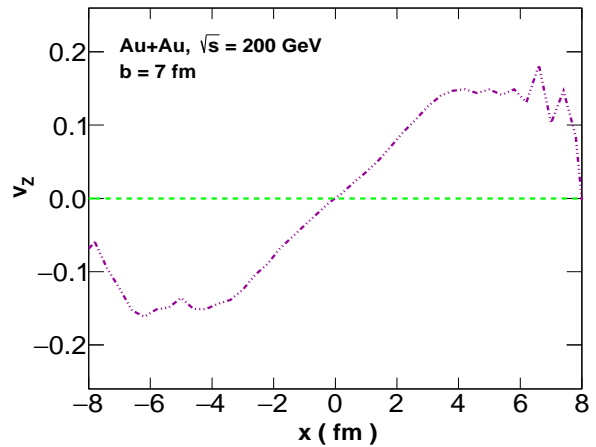}
\caption{\label{fig:vz_x} Longitudinal velocity profile along $x$ axis from the modified
initialization for a Au + Au collision at top RHIC energy $\sqrt{s_{NN}} = 200 \ GeV$
with impact parameter $b = 7 \ fm$.}
\end{figure}

In the BAMPS calculations performed before \cite{Uphoff:2014cba}, the initial condition was
the production of quarks and gluons 
according to the ``hard'' nucleon-nucleon binary collisions. The ``soft'' particle production
from the wounded participant nucleons, which is essential for the global vorticity,
was neglected.  When calculating the total angular momentum ${\bf J}$ by
\begin{equation}
{\bf J} = \sum\limits_{i} \, {\bf r}_i \times {\bf p}_i \,,
\end{equation}
where ${\bf r}_i$ and ${\bf p}_i$ are the position and momentum of $i$-th particle,
we find that the default initialization leads to zero orbital angular momentum because of
the mirror symmetry with regards to the $y-z$ plane [see Fig. \ref{fig:config} ($c$)].
To consider an orbital angular momentum in the present study, we will modify the default
initialization instead of adding the ``soft'' particle production. To be specific, we randomly
choose $1/5$ of the particles from the default initialization and treat them as if they were
produced from the wounded nucleons. To give these particles a global orbital angular
momentum, we randomly choose such particles at positive (negative) $x$ with negative 
(positive) $p_z$ and change the sign of $p_z$, so that the asymmetry between particles
with positive and negative $p_z$ is equal to $A_{part}({\bf x}_T, b)/5$, namely, 
\begin{equation}
\label{participant_asymmetry2}
\frac{dN_+/d{\bf x}_T-dN_-/d{\bf x}_T}
{dN/d{\bf x}_T}  = \frac{1}{5}A_{part}({\bf x}_T, b) \,,
\end{equation}
where $N_+$ ($N_-$) denotes the number of particles with positive (negative) $p_z$
and $N=N_++N_-$. Let $W$ be the probability for changing the sign of $p_z$ of
a certain particle. For instance, some particles at positive $x$ with negative $p_z$ will
change their sign to be positive. The average number of those particles is obviously
$\frac{1}{2}dN/d{\bf x}_T W$.  Thus, we have
$dN_-/d{\bf x}_T=\frac{1}{2}dN/d{\bf x}_T (1-W)$ and
$dN_+/d{\bf x}_T=\frac{1}{2}dN/d{\bf x}_T (1+W)$. Putting these relations in 
Eq. (\ref{participant_asymmetry2}) gives $W=A_{part}({\bf x}_T, b)/5$, with which
we modify the default initialization. The modification leads to $J_y=1.4\times10^4 \hbar$
for Au+Au collisions at top RHIC energy $\sqrt{s_{NN}} = 200 \ GeV$ with impact parameter
$b=7 \ fm$. This is $5-7$ times smaller than the result from AMPT model
 \cite{Jiang:2016woz} and the analytical result from \cite{Gao:2007bc}, and is 
 almost the same as the result from HIJING model \cite{Deng:2016gyh} and from
the recent work \cite{Oliva:2020doe}.
Figure \ref{fig:vz_x} shows the profile of the initial longitudinal velocities of sheets
transverse to $x$ axis, where $v_z=\sum_i p_{iz}/\sum_i E_i$. The sum is over
particles from all rapidity in a small $\Delta x$ window.

\begin{figure*}[t]
\includegraphics[width=8cm,height=6cm]{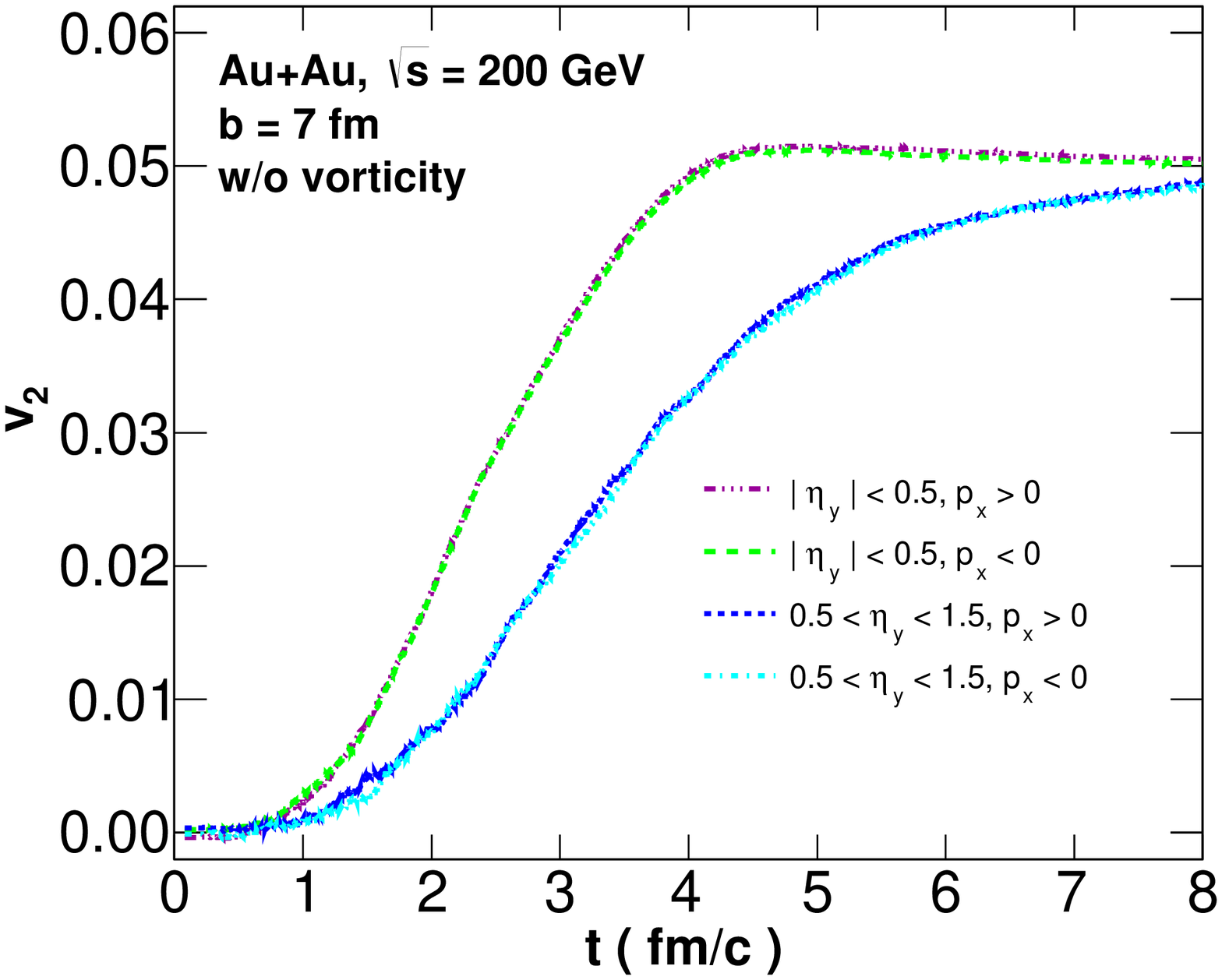}
\includegraphics[width=8cm,height=6cm]{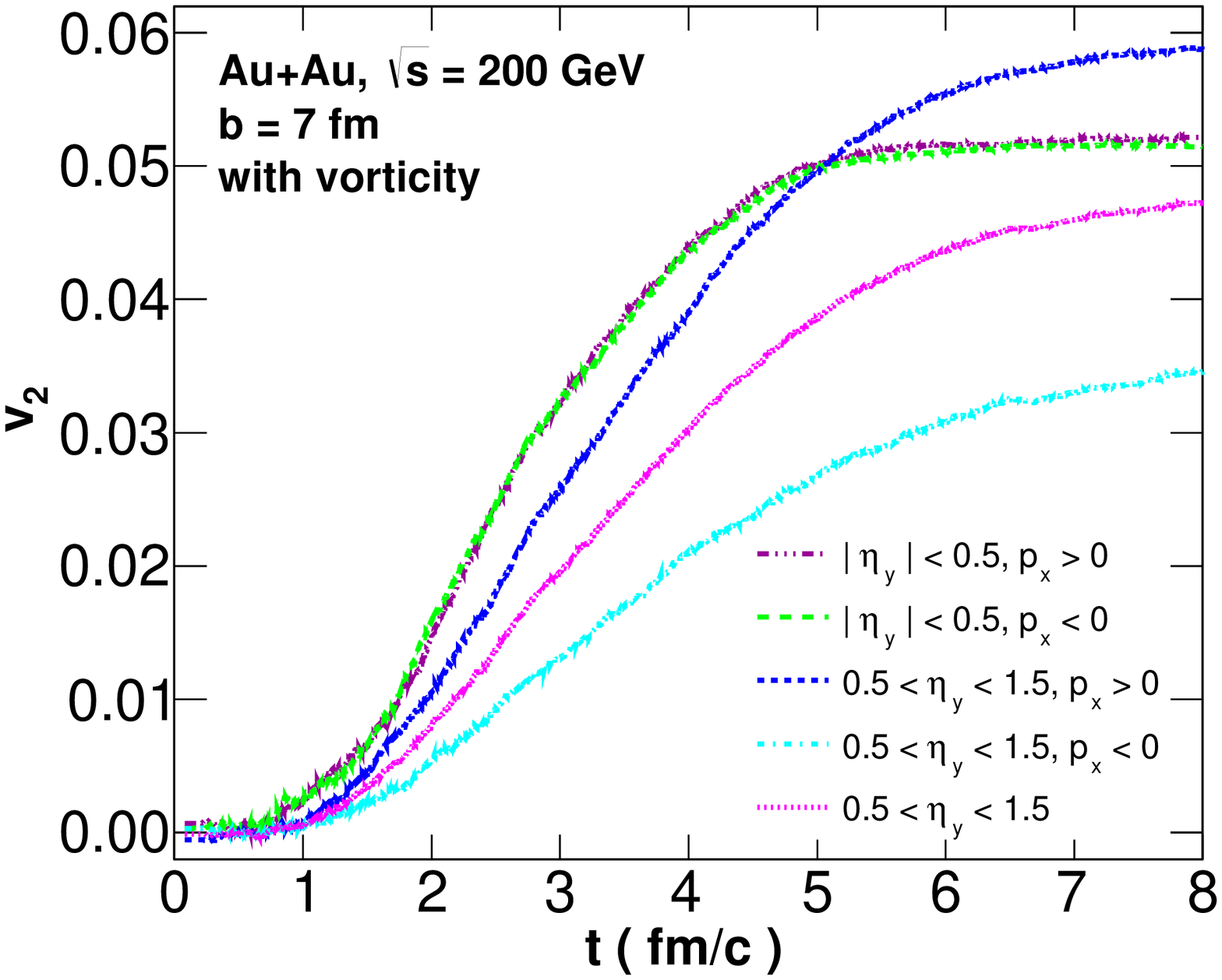}
\caption{\label{fig:v2_pQCD} 
Time evolution of the elliptic flow $v_2$ from BAMPS calculations for Au+Au collisions 
at $\sqrt{s_{NN}} = 200 \ GeV $ with $b = 7 \ fm$. $v_2$ is evaluated
for particles at mid-rapidity and higher rapidity with positive or negative $p_x$.
The left panel shows the results with the default initialization without a global vorticity, 
whereas the right panel shows the results with the modified initialization with a global vorticity.}
\end{figure*}

After the initialization, the space-time evolution and collisions of quarks and gluons
are calculated by BAMPS.  All $2 \leftrightarrow 2$ and $2 \leftrightarrow 3$ pQCD 
processes for gluons and $u$, $d$, and $s$ quarks are included. The details of 
the implementation can be found in \cite{Uphoff:2014cba}. We did some improvements
for the present study. The test particle number $N_{test}$ is increased from $250$
to $25000$ to get higher statistics. The cell length in the transverse plane is reduced 
from $dx=dy=0.4 \ fm$ to $0.05 \ fm$. The cell length in the space-time rapidity is
about $d\eta_s \approx 0.02$ and almost unchanged.
The local quantities such as the particle number and energy
density, Debye mass, and the mean-free path etc., which are needed to determine
the transition cross-sections and the freeze-out condition, are calculated locally
in a small box with $dx=dy=0.4 \ fm$ and $d\eta_s \approx 0.02$.  
Other setups are the same as made in \cite{Uphoff:2014cba}.
 
In the following we show the numerical results from BAMPS calculations for
Au+Au collisions at top RHIC energy $\sqrt{s_{NN}} = 200 \ GeV$ with 
$b = 7\  fm$. The initial condition is either the default initialization with zero ${\bf J}$ or
the modified one with non-zero ${\bf J}$, in order to demonstrate the significance of
a global  vorticity.  

We calculate the elliptic flow parameter $v_2$ in two rapidity windows. Here the momentum
rapidity is defined as $\eta_y = 1/2\ln[(E + p_z )/(E - p_z )]$. The two rapidity windonws are
chosen at mid-rapidity $[-0.5:0.5]$ and higher rapidity $[0.5:1.5]$. $v_2$ is evaluated
by $v_2= \sum_i (p_{ix}^2-p_{iy}^2)/(p_{ix}^2+p_{iy}^2)$, where the sum is over the particles
with positive $p_x$, or the particles with negative $p_x$, or all the particles in the given
rapidity window, respectively. 

\begin{figure*}[t]
\includegraphics[width=8cm,height=6cm]{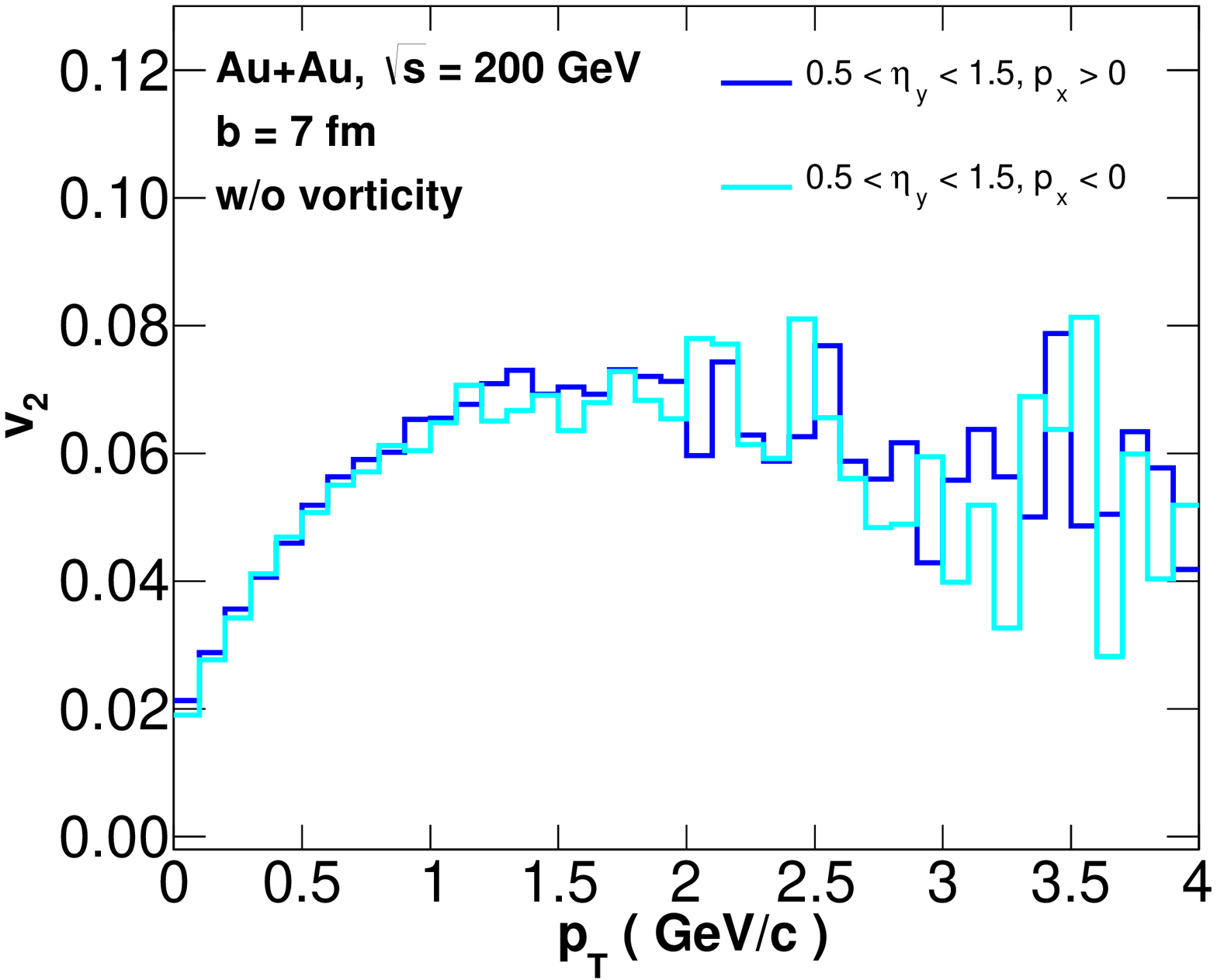}
\includegraphics[width=8cm,height=6cm]{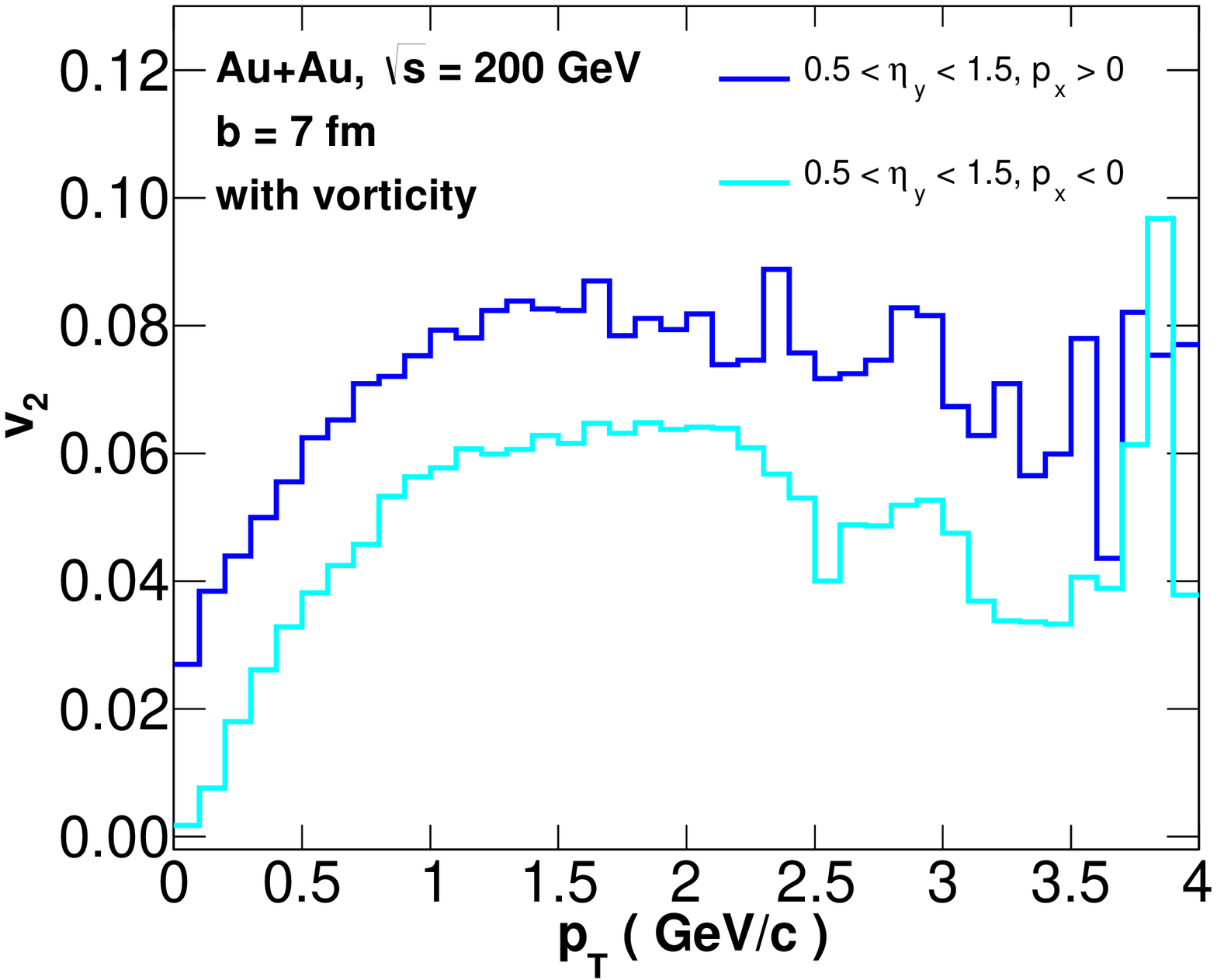}
\caption{\label{fig:v2_pt} Same as Fig. \ref{fig:v2_pQCD}, but for the $p_T$ dependence
of the final $v_2$ in the rapidity window $[0.5:1.5]$.}
\end{figure*}

Figure \ref{fig:v2_pQCD} shows the buildup of $v_2$. The left panel depicts the results
with the default initialization, whereas the right panel depicts the results with the modified
initialization. We see no difference between $v_2$ for particles with positive
$p_x$ and that for particles with negative $p_x$ in both rapidity windows for the default
initialization without a global vorticity, as it should be due to the mirror symmetry with
regards to the $y-z$ plane, see Fig. \ref{fig:config} ($c$). The mirror symmetry is broken
for the modified initialization with a global vorticity, see Fig. \ref{fig:config} ($d$).
However, the symmetry between the distribution at rapidity $\eta_y$ and that at $-\eta_y$
by changing $p_x$ to $-p_x$ holds. This is the reason that $v_2$ of particles with positive
$p_x$ in the mid-rapidity window $[-0.5:0.5]$ is the same as that for particles with
negative $p_x$ for the modified initialization. It is not the case for $v_2$ in the higher 
rapidity window $[0.5:1.5]$, because the Bjorken boost invariance is not valid any more
due to the global vorticity. We see that at the higher positive rapidity the $v_2$ of particles
with positive $p_x$ is much larger than that of particles with negative $p_x$ for the modified
initialization, although the $v_2$ for all particles, depicted by the red curve in the right
panel of  Fig. \ref{fig:v2_pQCD}, is only a little smaller than that for the default initialization.
We call this new finding the splitting of the elliptic flow at non-zero rapidity in presence of 
a global vorticity. The splitting is also seen in the $p_T$ dependence of $v_2$ at the final
time, as shown in the right panel of Fig.\ref{fig:v2_pt}. When comparing the results with and
without the vorticity in Fig. \ref{fig:v2_pt}, we see that the vortical motion shifts the curve
of $p_x > 0$ up and shifts the one of $p_x < 0$ down. Note that the hadronization and 
the hadronic interactions are not included in the present BAMPS calculations.
The hadronic $p_T$ dependence would be higher at intermediate $p_T$ due to the quark recombination. Nevertheless, the splitting effect would
remain in a full transport calculation.

As mentioned before and illustrated in Fig. \ref{fig:config} $(d)$, the global vorticity 
seems to ``drive'' the system outwards in quadrant $I$ and to ``press'' the system inwards
in quadrant $IV$.
Therefore, qualitatively, the vorticity will enhance the transverse flow in quadrant $I$ and reduce
the transverse flow in quadrant $IV$. In other words, the vorticity will cause an additional flow 
in quadrant $I$ and an antiflow in quadrant $IV$. This is the reason why at positive 
rapidity the $v_2$ of particles with positive $p_x$ is always larger that that of particles
with negative $p_x$.

We note that the initial fluctuation was not taken into account in the initial particle distribution
for BAMPS calculations. The pure initial fluctuation may also lead to the splitting of the
elliptic flow.  Whether $v_2$ of particles with positive $p_x$ is larger or smaller than that
of particles with negative $p_x$, is random on the event by event basis.
This is different from the vorticity induced splitting.

The sizable splitting of the elliptic flow, shown in Figs. \ref{fig:v2_pQCD} and \ref{fig:v2_pt}
from our calculations, serves as a demonstration of the significance of the global vorticity. 
When decreasing the number of the ``soft'' particles in the initialization to be $1/10$
of all particles, (instead of $1/5$ used), the global orbital angular momentum is decreased
to $J_y=0.63\times10^4 \hbar$ and the initial profile of the longitudinal velocity becomes
less steep than that shown in Fig. \ref{fig:vz_x}. For this initial condition, the size of the splitting
of the elliptic flow at the final time is about $0.014$, which is smaller than the value of 
$0.024$ taken from Fig. \ref{fig:v2_pQCD}.  Qualitatively, the larger the global
orbital angular momentum, and/or the more pronounced the initial vortical configuration,
and/or the smaller the shear viscosity, the more significant is the splitting of the elliptic flow.
To make these dependence clear, further theoretical investigations are needed.
From the experimental side, a confirmation of the splitting of the elliptic flow by
experimental measurements
could prove the existence of a global vorticity and the measured size of the splitting 
could be used to constrain the vortical configuration in the early stage of heavy-ion collisions.

In summary, we have studied the possible experimental significance of a global vorticity
in non-central relativistic heavy-ion collisions and proposed the splitting of the elliptic flow
at non-zero rapidity as a new measurable observable. The idea is repeated here. 
The collective motion of a global vorticity will break the mirror symmetry with regards to
the $y-z$ plane and the Bjorken boost invariance, which hold in the case without a global
vorticity. The difference in the motion above and under the $y-z$ plane in the coordinate
space can be transferred into the momentum space,which makes a splitting of the elliptic
flow parameter $v_2$ at non-zero rapidity with regards to the sign of $p_x$.
By employing the parton cascade model BAMPS for the default and modified initialization
of quarks and gluons in a Au + Au collision at RHIC energy $\sqrt{s_{NN}} = 200 \ GeV$
with impact parameter $b = 7 \ fm$, we obtained a significant splitting of the elliptic flow
$v_2$ in the rapidity window $[0.5:1.5]$, when a global vorticity is present.

We have to note that since the particle distribution function is given in Fourier-series,
\begin{equation}
\frac{dN}{d^2p_T d\eta_y}=\frac{dN}{2\pi p_T dp_T d\eta_y} 
\{ 1+2\sum_n v_n \cos[n(\psi-\Psi_n)]\} \,,
\end{equation}
the splitting of $v_2$, $\Delta v_2$, can be mathematically expressed by all $v_n$ and
$\Psi_n$. In particular, all even $v_n$ and $\Psi_n$ disappear. If higher order flow
parameters can be neglected, we have approximately 
$\Delta v_2 \approx(8/3\pi)v_1 \cos(\Psi_2-\Psi_1)$. 
$v_1$ and $\Psi_2-\Psi_1$ fluctuate event-by-event.
We see that $\Delta v_2$ and $v_1$ have the same sign. Qualitative analysis as well as
numerical calculations presented in this study show that the defined $\Delta v_2$
at positive rapidity is positive. However, the measured $v_1$ 
\cite{PHOBOS:2005ylx,STAR:2005btp,STAR:2008jgm,STAR:2014clz}
is negative at positive rapidity. This may be due to the
contribution of non-flow effects to the particle distribution function. It seems that $v_1$ is not
completely equivalent to $\Delta v_2$ that we proposed. Since non-flow effects have
been already handled in data analyses for $v_2$ measurements, we suggest
the observation of the splitting effect of the elliptic flow in experiments.

The authors would like to thank Pengfei Zhuang and Dariusz Miskowiec for fruitful discussions. 
This work was financially supported by the National Natural Science
Foundation of China under Grants No. 11890710, No. 11890712, and No. 12035006.
C.G. acknowledges support
by the Deutsche Forschungsgemeinschaft (DFG) through the grant
CRC-TR 211 ``Strong-interaction matter under extreme conditions''.
The BAMPS simulations were performed at Tsinghua National Laboratory
for Information Science and Technology and on TianHe-1(A) at National
Supercomputer Center in Tianjin. 

\nocite{*}

\bibliography{References}

\end{document}